# Crowdsourcing for Usability Testing

Di Liu, Matthew Lease, Rebecca Kuipers, and Randolph Bias
School of Information
University of Texas at Austin
{ericliu, ml, rkuipers, rbias}@ischool.utexas.edu

## ABSTRACT
While usability evaluation is critical to designing usable websites, traditional usability testing can be both expensive and time consuming. The advent of crowdsourcing platforms such as Amazon Mechanical Turk and CrowdFlower offer an intriguing new avenue for performing remote usability testing with potentially many users, quick turn-around, and significant cost savings. To investigate the potential of such crowdsourced usability testing, we conducted two similar (though not completely parallel) usability studies which evaluated a graduate school's website: one via a traditional usability lab setting, and the other using crowdsourcing. While we find crowdsourcing exhibits some notable limitations in comparison to the traditional lab environment, its applicability and value for usability testing is clearly evidenced. We discuss both methodological differences for crowdsourced usability testing, as well as empirical contrasts to results from more traditional, face-to-face usability testing.

## Categories and Subject Descriptors
H.5.2 [**Information Interfaces and Presentation**]: User Interfaces – *Evaluation/methodology*; H.5.2 [**Information Interfaces and Presentation**]: User Interfaces – *User-centered design*; H.1.2 [**Models and Principles**]: User/Machine Systems – *Human Factors.*

## 1. INTRODUCTION
Usability has been defined as *the extent to which a product can be used by specified users to achieve specified goals with effectiveness, efficiency and satisfaction in a specified context of use* [15]. Usability tests should be done early and often since *usability plays a role in each stage of the design process* [22]. Unfortunately, the cost and effort required to recruit and test participants, engage observers, and purchase or rent equipment can be prohibitive. Thus while usability testing is important to the success of any website, cost and delay of user feedback is often deemed too expensive in practice for ongoing website development and maintenance. However, a recent move towards a cost-benefit analysis approach to usability has led to increased attention to return-on-investment (ROI) when considering incorporating usability evaluations into a product/site development effort [4]. The cost of usability tests suggests a careful tradeoff to be balanced in allocating limited resources between usability evaluation and design [27].

In this paper, we investigate an alternative way to perform usability tests: crowdsourcing, a relatively new and quickly growing phenomenon. One very prominent example, Amazon's Mechanical Turk (mTurk, [mturk.com](mturk.com)), provides a commercial marketplace for so-called "Human Intelligence Tasks" (HITs). Employers have access to a diverse, on-demand, scalable workforce, and workers have a constant and diverse supply of thousands of tasks to select from whenever they choose to work. Another vendor, CrowdFlower ([crowdflower.com](crowdflower.com)), provides value-added service atop multiple crowdsourcing "channels", such as mTurk and SamaSource ([samasource.org](samasource.org)).

To investigate the potential of such crowdsourced usability testing, we conducted two usability studies evaluating a graduate school's website: one via a traditional usability lab setting, and the other using mTurk and CrowdFlower. While each test was utilized individually to inform a redesign of the website, comparing the two tests in tandem enabled us to assess the quality of the findings of each method and to characterize situations and tasks for which crowdsourced usability testing is viable. Because each test involved different user populations (global workers vs. university students), a limitation of our study is some variation between tasks each group performed. However, this difference in population demographics of each group represents a fairly inherent difference between the methods themselves, an issue we expand upon in subsequent discussion.

We consider the following research questions in this work. *How well can usability tests be performed on crowdsourcing platforms? What kinds of tasks would be best for crowdsourcing usability tests? How valuable are the crowdsourcing usability test results compared to traditional lab usability tests? How might we design a better crowdsourcing usability test?*

The remainder of this paper is organized as follows. First, we discuss prior work on crowdsourcing usability tests. We then describe our experimental method for both lab and crowdsourcing usability tests. This is followed by a comparison and discussion of the results of those two tests. Finally we present recommendations on both the use of crowdsourcing for usability testing and the proposed directions for future research.

## 2. RELATED WORK
Usability evaluation has enjoyed a rich history in the last four decades, There has been an evolution from end-user testing in the lab [26], to inspection methods [23], to remote testing [3], all in the interest of finding thorough, reliable, efficient methods of collecting user data to inform and validate user interface designs.

While usability engineering has enjoyed a rich history and robust growth, there has been precious little empirical study of the validity and efficiency of various usability engineering methods. Empirical studies that have been undertaken have tended not to compare multiple methods, but rather to compare multiple practitioners/teams carrying out the same method, such as in the Comparative Usability Evaluation (CUE) studies carried out by Molich and Dumas [21]. In contrast, we report an explicit, empirical comparison of two methods, traditional, face-to-face usability lab testing, and crowdsourced usability testing.

**Crowdsourcing**

*Crowdsourcing is the act of taking a job traditionally performed by a designated agent and outsourcing it to an undefined, generally large group of people in the form of an open call* [10]. People perform crowdsourced work for various reasons: payment, altruism, enjoyment, reputation, socialization, etc. [24]. Crowdsourcing is becoming increasingly popular and has been studied as a usability engineering method [17]. With crowdsourced usability testing, one can tap into a wide diversity of users to test an online website or application.

While crowdsourcing boasts various strengths vs. prior practices, various problems currently limit the potential of existing platforms like mTurk. Worker anonymity, coupled with lack of sufficient accountability and task-based payment entices some workers to complete many tasks poorly, or even utilize bots (contrary to terms of service). For example, "spammers" or cheaters may try to maximize their individual profits without care as to the quality of work they perform. They might answer questions randomly [8], jeopardizing the validity of study results based upon their answers. While participants who do not fully engage in the traditional usability test in lab settings also exist, they have not been nearly so prevalent as in crowdsourcing today. Kittur, Chi, and Suh [17] thus recommend contrary to traditional usability design to make crowdsourced tasks *more effortful* to complete such that it is no easier to cheat than to do complete the task correctly. Another challenge they identified is potentially low ecological validity: the experimenter has little control of the setting in which the mTurk user carries out a task.

CrowdFlower allows customers to upload tasks to be carried out on mTurk or other crowdsourcing "channels". It takes large, data-heavy projects and breaks them into small tasks that can be done by crowd workers. Results are then aggregated with higher-level controls for quality using "Gold Units": hidden tests randomly distributed through the tasks that a worker completes. These tests have known answers, facilitating easy evaluation of a worker's output. If a worker makes too many mistakes on Gold Units, his/her answers will be automatically rejected by the system, thereby simplifying quality management for customers.

Other crowdsourcing platforms like oDesk (odesk.com) [5] or the internal system described by Freebase (freebase.com) [18] adopt a different approach: paying workers more hourly wages rather than by volume of work completed. In this way it is expected that workers will produce higher quality work because there is no benefit to rush, while to the contrary there is an incentive to do good work to maintain continuing employment. On the other hand, work may be completed more slowly since there is no explicit financial incentive for quick task completion. Workers could potentially stretch out their hours in this fashion (though oDesk also provides tools for worker monitoring to verify remove workers are actively engaged when "clocked-in"). Usability tests can be seriously affected by such timing issues because time-on-task is a commonly utilized usability metric.

Crowdworkers come from all over the world. On mTurk, Amazon boasts more than 500,000 Mechanical Turk Workers in 190 countries, while independently collected self-reported demographics from workers indicate primarily U.S. and India origins [25]. Reported worker ages range from teenagers to senior citizens, with education levels ranging from high school to doctoral degrees. Some workers depend on income from mTurk for a living, while many just earn a few extra bucks while passing the time [12]. While the question of fair pay for globally distributed crowd workers is notoriously difficult to determine [20], in terms of effect, more pay can be expected to attract more workers, including more spammers. Greater financial incentives have been seen to increase quantity but not quality of work [19].

**uTest**

While Amazon and CrowdFlower have significant market share in the micro-task market segment (e.g. tagging and labeling), uTest (utest.com) is the only crowdsourcing company we are aware of that has specifically targeted usability testing customers. While we do not utilize or evaluate uTest in this work (nor have we received any support from them), their model and workflow for crowdsourced usability testing is sufficiently interesting to merit briefly summarizing and comparing to the approaches we do utilize and evaluate (traditional lab-based testing vs. crowdsourced testing via mTurk and CrowdFlower).

The uTest community is comprised of a broad group of testers spanning many locations, languages, operating systems (OS), browsers, and devices. uTest customers specify test requirements such as geographic location, OS, and browsers, then upload testing scripts. uTest proceeds to identify and invite qualified testers from its large community. Testers who accept the invitation then test the website/app's functionality or usability according to the provided testing script. Finally, requestors approve or reject each tester's report based on quality [28][29].

Controlling demographics on mTurk is more difficult than with uTest. Since mTurk workers are anonymous, requesters must test worker qualifications or rely on self-reported demographics. In contrast, uTest requires testers to provide demographic information during registration. While this enables uTest to more easily support targeted demographic testing, available crowdworkers likely do not cover all demographics of potential interest, such as users with limited prior internet experience, etc.

While uTest offers relatively easy usability testing, their cost model is relatively higher than other crowdsourcing platforms. For example, while an hourly rate of $1-$2 is more typical on mTurk, uTest prices are typically $25 or more per test per participant [25][30] (though this premium likely provides greater quality in return). While both vendors offer potentially significant cost savings vs. traditional lab tests, one must be sure to account for other incident costs when comparing alternatives.

## 3. METHODS

Our study involved conducting two usability tests of a graduate school's website: one performed via a traditional lab usability test, while the other was conducted via crowdsourcing.

Each test asked participants to perform a set of tasks on the website. The two groups of participants performed similar but distinct tasks reflecting the different user populations involved.

Our interest in originally conducting the usability tests was to evaluate the school's website for current students, so we recruited current students familiar with the site as participants. However, none of our students were users on the crowdsourcing platform; we return to this limitation of our comparison of tests in later discussion. Consequently, while participants in the lab usability test performed tasks designed for students with experience using the website, the crowdsourcing usability tasks were designed for prospective students who had never used the website before. Participants in each group (subgroup) all performed the same tasks as the others in their subgroup.

Usability test objectives in both tests were as follows:

1. Determine design inconsistencies and usability problem areas within the user interface and content areas. Potential sources of error may include:
   a. *Navigation errors*: failure to locate functions, excessive keystrokes to complete a function, failure to follow recommended screen flow
   b. *Presentation errors*: failure to locate and properly act on desired information in screens, or selection errors due to label ambiguity
   c. *Control usage errors*: improper toolbar or entry field usage.
2. Exercise the application or website under controlled test conditions with representative users. Data assessed whether usability goals for an effective, efficient, and well-received user interface were achieved.
3. Establish baseline user performance and user-satisfaction levels of the user interface for future usability evaluations.

## 3.1 Test One: Traditional Lab Usability Test

The lab usability test was performed in a traditional usability lab setting at the graduate school with five participants. All participants in this part of the test were current students from the school, all regular users of the website, and all volunteers. A Dell laptop computer with Mozilla Firefox and HyperCam3 [1] installed on it was used. Participants' interaction with the website was monitored by two testers: a silent observer and a facilitator, who tested and interacted with each participant. Test sessions were recorded by HyperCam3 for later analysis.

During the usability test, participants were first introduced to the goals and method of the test. They then completed a pre-test demographic and background information questionnaire. Participants were asked to perform five tasks related to everyday use of the website by current students:

*Task 1*. You sometimes have extra time in your schedule. Please find the list of extracurricular workshops that the school offers.

*Task 2*. You know that at the end of every semester there is an open-house where students display posters of their work. Find the information about the school's open house.

*Task 3*. You are interested in taking a course on Usability next semester. Who will be teaching Usability in the Fall, 2011 semester? Where and when will this class meet?

*Task 4*. You want to get a student job helping a professor on a research project involving archives. What projects are being done and what professors should you approach?

*Task 5*. A lot of your classes deal with technology that you are unfamiliar with. What assistance is available?

Each task was considered to be completed when the participant indicated that either the goal had been achieved or that he/she would normally stop using the website to achieve the goal. Participants were asked to verbalize their thought process as they worked on each task. Narrating the process necessarily increased the time spent on each task, however it provided explanations of why certain tasks were or were not difficult to perform and what design features aided or hindered a given process. As regular users of the website, the participants would also mention problems they had experienced during previous interactions with the website. In their stream-of-consciousness narration, the participants also sometimes mentioned minor frustrations that they did not always remember in any detail later on.

After all five task scenarios were attempted, participants were asked six qualitative questions regarding their experiences:

*1. Is there anything in particular that you would like to tell us about the website or your experience with it?*

*2. What are your thoughts on the structure of the website?*

*3. What are your thoughts on the aesthetics of the website?*

*4. Did you have any suggestions for improving the website?*

*5. Was there anything you particularly liked and would not want to see changed?*

*6. Was there anything you particularly disliked and would like to have removed or modified?*

These questions allowed the participants to synthesize their own thoughts regarding the website, looking at both their recent experience during the test and their long-term experience as regular users of the website. The questions also helped ally the participants with the usability testers to provide the best quality of feedback. As current users of the website, participants were naturally invested in the improvement of the website and were given the opportunity to assist in its (ongoing) development.

Tests took from 20-40 minutes from greeting to goodbye.

## 3.2 Test Two: Crowdsourcing Usability Test

The crowdsourcing usability test was done in two rounds. Because we had no prior experience conducting a crowdsourced usability evaluation, we first ran a pilot test with 11 participants. After analyzing results from the pilot test, we then modified the test and ran a final test with 44 additional participants.

The crowdsourced usability test was performed on the mTurk crowdsourcing platform. The pilot test participants were recruited directly from mTurk. The final test participants were also recruited from mTurk but via CrowdFlower as intermediary. Participants performed the tests in their own environments. Their actions on the website were not recorded.

Participants were first directed to a survey and asked to fill out a demographic questionnaire. They were then asked to perform a series of four tasks, discussed further below, on the website. These tasks represent what new users of the website would be doing on the website. After the tasks were completed, the participants were asked to answer a series of open-ended questions regarding their experience with the website. None of these participants indicated any prior experience with the website being tested. Unlike in the traditional lab test, participants in this test received compensation for participating.

---

[1] HyperCam3 is a screen capture software developed jointly by Solveig Multimedia and Hyperionics LLC.

# 4. CROWDSOURCED USABILITY TEST

Designing crowdsourced usability testing is different from designing traditional lab usability testing. First, since test facilitators do not interact directly with participants, instructions and tasks in the crowdsourced usability testing must be described specifically and totally unambiguously – there is no chance to offer subsequent clarification. Second, since crowdsourcing test participants are likely to be less engaged in the goal of the test and have a higher chance (compared to lab testing) of not really trying, the survey used in the crowdsourcing usability test should be designed to make it difficult for the user to cheat [17].

In order to ensure high quality data, our survey was designed under the following methods introduced by Kapelner and Chandler [16]. First, the perceived value of the survey was increased by informing the participants that the results they provided would be used in an academic study. Second, instead of multiple choice questions for which workers can randomly select answers, we used blank-filling questions which require users to go to the website and look for information in order to find the answer and continue. Since we did this, the participants were forced to slow down and spend time on the task. Third, to get high quality feedback, we indicated in instructions that workers who gave substantial feedback would be given a bonus of up to 100%, while those giving random answers would be rejected.

## 4.1 Pilot Test

The pilot test crowdsourced usability test was performed by 11 participants recruited directly from mTurk. We first requested they supply demographic information including age, gender, and highest level of education attained. We then directed them to the website, asking them to complete a series of tasks and answer a set of open-ended questions regarding their experience.

The whole survey was designed to be done in 10 minutes. We offered $0.15 for each HIT. The pilot test was launched on a Sunday afternoon, with results from all 11 participants available in under three hours. The total cost including bonuses given to participants who did a good job and mTurk commission was $2.92 ($1.10 as bonuses and $0.17 as mTurk commission).

### 4.1.1 Pilot Test Results

The results of the pilot test informed our subsequent test design. The average time spent on the HIT was approximately 13 minutes which was longer than expected. The fastest worker used only 7 minutes to complete the HIT. Workers with a Bachelor's degree or higher education level completed the HIT significantly faster than those with associate degree or lower. Workers with higher education levels were likely more familiar with educational institution websites like the one being tested.

In the open-ended question section, it was clear that users were not interested in giving detailed feedback to open-ended questions like "*Please give some feedback regarding the website.*" Answers submitted to such questions were all one sentence or less, such as "*very interesting*", "*nice website*". Given the instructions we had provided, participant responses suggested they believed such brevity would still earn them payment while enabling them to complete the HIT quickly. All workers were deemed to have completed the HIT as instructed, all results were accepted, and no spammers were identified.

### 4.1.2 Test Redesign

In response to the problems identified from the pilot test, the usability test was redesigned. First, an additional demographic question "*Do you have any previous experience of applying for any kind of college or grad school*" was added to separate experienced workers from inexperienced workers. Tasks in the later version of the survey were also stated more clearly to avoid misunderstandings or multiple correct answers to a question. In order to get more substantial feedback, the open-ended questions were broken down into more detailed questions such as "*What are your thoughts regarding the structure of the website?*" and "*Is there anything about the website you particularly liked?*" Because the pilot test took the participants longer than expected to finish, we also raised the compensation for the final test to provide appropriate pay to attract and engage workers. Finally, we decided to recruit mTurk participants for the final test via CrowdFlower, whose value-added service suggested potential for getting faster respond speed and higher quality results.

## 4.2 Final Test

The final crowdsourced usability test was completed by 44 mTurk participants. We required workers' to self-report demographic information including age, gender, and highest level of education attained. Participants were then directed to the school's website and asked to complete a series of tasks:

*Task 1.* Assume you want to apply for a Master's degree in this graduate school. What is the minimum GPA requirement?

*Task 2.* How many semester hours of course work must a student take to earn the Master's degree in the school?

*Task 3.* Please find the link to the list of faculty specializations in the school and paste the link below.

*Task 4.* Assume you need financial aid to help you attend the school. What funding resources are available? Please find the webpage and paste the link below.

Participants in the final test were also asked to record and report time (in minutes) spent on each task. After finishing the tasks, workers were asked a set of open-ended post-test questions:

*1. What are your thoughts on the structure of the website?*

*2. What are your thoughts on the aesthetics of the website?*

*3. Did you particularly like anything about the website?*

*4. Did you particularly dislike anything about the website?*

*5. Would you like to mention anything else about it?*

*6. If you wanted to earn a degree in an information school, how likely would you apply to this graduate school based on the experience on its website? (1-7 scale where 1 stands for "not likely at all" and 7 stands for "very likely")*

The entire survey was designed to take about 15-20 minutes for each participant. We offered $0.40 for each HIT in the final test. The final test was launched on a Sunday afternoon. The results of all 44 participants came back in less than an hour. The total cost of the final test including CrowdFlower commission was 44 * $0.40 * 1.33 = $23.41.

### 4.2.1 Final Test Results

The results for the final test came back even more quickly than in the pilot mTurk study. We hypothesize that HITs posted via CrowdFlower have a good reputation among crowd workers and many workers are searching for HITs posted by them.

In the final test, the majority of the workers completed the test as instructed. However, approximately 30% of the workers (14 out of 44) were manually flagged as spammers. This is because the answers they provided appeared to be random. For example, some gave nonsensical answers like "7" for Task 1 and the URL "www.schools.org/specializations" for Task 3. Though mistakes made by test participants in a usability test are generally due to the usability problem of a website, it seemed that these spammers did not even go to the graduate school's website through the link we provided as instructed.

Unfortunately, on CrowdFlower, unlike mTurk, there is no way to reject poor work once completed. Consequently, use of "Gold Units" to screen workers is critical. The challenge with usability testing, of course, is that any mistake can be legitimate due to usability problems of the website. However, one kind of possible Gold Unit for usability might be to simply ask workers to report the first word on a given website. Such Gold Units could minimally verify that workers went to the website. However, such Gold Units would not not verify the workers really tried to do the usability tasks well. Given the nature of usability testing, this makes it challenging to tell if the worker is cheating or not by the mistakes made. We screened out spammers by manually checking their answers to see they were at all reasonable.

Times-on-task reported by some workers were suspect. For example, one worker reported that he/she spent 5, 8, 5, and 15 minutes on each of the four tasks, yet he/she finished the whole survey in 23 minutes. We expected that workers would not be timing themselves with extreme accuracy, but the ultimate results were too inaccurate to usefully employ time-on-task data in the evaluation of the web site.

By separating open-ended questions into more detailed ones, the feedback submitted was much better than in the pilot test. Even though most workers still gave only one-sentence answers to each question, they had to give at least four sentences to complete all the questions. Quality of the feedback was higher than in the pilot test. This increase was likely due to workers better understanding the more specific final test questions.

## 5. DISCUSSION

The results of the two studies had notable similarities as well as differences. Lab usability testing and crowdsourced usability testing were performed with different numbers of participants; the demographics were different; the times spent on tests were different; the specific tasks were different; the monetary costs were also different (Table 1).

The time spent by participants in the crowdsourced usability test was significantly less than the time spent by participants in the lab usability test. In the lab usability tests, it took approximately 30 minutes for each participant to perform the test. It also took time for the participant and test facilitator to schedule the test.

**Table 1. Comparison between lab usability test and crowdsourced usability test**

|  | Lab Usability Test | Crowdsourced Usability Test |
|---|---|---|
| **Participants** | 5 | 55 (14 spammers) |
| **Participant Demographics** | Students | Crowdworkers |
| Age | 24 to 33 | 19 to 51 |
| Education level | Bachelor's degree and Master's degree | All levels |
| Experience with similar websites | Yes: 100% | Yes: 77% No: 23% |
| **Speed** | Approximately 30 min. per session. | Less than 4 hours total. |
| **Participant Costs** | None | $2.92 for pilot test $23.41 for final test (Avg: $0.48/tester) |

**Table 2. Usability problems found from lab usability test and crowdsourced usability test**

| Major Problems Identified | Lab Usability Test | Crowdsourcing Usability Test |
|---|---|---|
| Font size too small | ✓ | ✓ |
| Out-of-date information | ✓ | ✓ |
| Menu overlap | ✓ | ✓ |
| Irrelevant picture | ✓ | ✓ |
| Invisible tools | ✓ |  |
| Information not cross-linked | ✓ |  |
| Lack of sort function | ✓ |  |
| Navigation unclear |  | ✓ |
| Search box difficult to locate |  | ✓ |

**Table 3. Advantages and Disadvantages of crowdsourced usability tests over lab usability tests**

| Advantages | Disadvantages |
|---|---|
| More Participants | Lower Quality Feedback |
| High Speed | Less Interaction |
| Low Cost | Spammers |
| Various Backgrounds | Less Focused User Groups |

Crowdsourced usability test participants had a wide variety of backgrounds. From the demographic questionnaire results, the participants' age ranged from 19 to 51. Most of them (68%) had a Bachelor's degree, but there were also workers with an Associate degree, Master's degree or Doctor's degree. In comparison, participants of the lab usability test ranged in age from 24 to 33 and all had at least some graduate-level education.

The usability problems of the website identified by both groups overlapped significantly despite their differences (Table 2). Major problems such as menu overlap and irrelevant pictures were identified by both lab test participants and crowd workers.

Lab usability test participants and crowdsourced usability test participants each identified problems that the other group did not. For example, lab test participants identified the lack of sort function in some pages, while crowd workers identified difficulty in finding the search box. The identification of different problems could easily be explained by the different tasks each group performed and their relative familiarity with the website.

Another issue we encountered was that in the lab tests, whenever a task or a question was not sufficiently clear, participants could ask for more instructions. With crowdsourced tests, in contrast, workers could not request any type of clarification in such circumstances (only via email, which never happened). When uncertain, crowd workers must therefore act upon their best guess, which may be wrong. Task design for usability tests, especially those to be done in crowdsourced usability tests, must be specific and unambiguous.

Similar issues appear in interpreting feedback from test participants. In lab tests, we can always ask participants for more details if they say something like "*The navigation menus are confusing.*" In crowdsourcing tests, it is more difficult (though not impossible) to send workers follow-up questions for explanations of what they meant by a given response. Unclear feedback is clearly less helpful than more specific feedback.

### 5.1.1 Advantages and Disadvantages
Overall, we found both advantages and disadvantages for crowdsourced usability testing (see Table 3).

**Advantages.** Recruiting participants from crowdsourcing platforms is much easier than asking people to come to the lab to perform a usability test. So it is easier to obtain more data from crowdsourcing usability tests.

Lab usability tests usually take about an hour per session. They cannot be done in parallel with each other (unless there are multiple lab spaces and multiple testers). The whole process might take days or even weeks to be done. Crowdsourcing usability testing saves travel, greeting time, and setting up processes. They can be done simultaneously so the whole process can be completed within hours.

The potential cost savings for crowdsourced usability tests are significant as well. While we used unpaid student volunteers for the traditional lab usability test, lab usability tests typically entail paying a participant for a one-hour session with a sum larger than their hourly wage. In comparison, the hourly rate for crowd workers is typically about $1.25. Of course, the total cost for usability tests is not only compensation to crowd workers. Time and monetary costs for test facilitators, labs, equipment, and travel can all be potentially lower in crowdsourced tests.

Because the time and monetary cost of crowdsourced usability tests is relatively low, it can be more affordably iterated. When a usability test is first designed, it can be run as a pilot test to see if there is any problem with the test itself. It can then be improved before being launched to more participants just as we did in this study. Crowdsourced usability testing may also be easier to be run throughout the development and maintaining process of a website because of its high speed and low cost.

Because crowd workers participate from all around the world, it is remarkably easy to conduct a test with participants from various backgrounds. This is especially beneficial to websites whose users are all around the world. Indeed, it would be an easy matter to launch parallel, crowd-sourced usability tests, each with a different user audience specified. In lab settings, the time and monetary costs rise significantly if companies want to test participants from other locations. While remote usability testing is possible [3], the set-up and test times are still additive.

**Disadvantages.** The quantity of feedback from a single crowdsourcing participant is much lower than the quantity of feedback from a single lab test participant. Many crowd workers seem to just want to get the HITs done as fast as possible in order to get paid with little care as to quality.

As such, the quality of usability test results from crowdsourcing is noticeably lower than those from the lab tests. Workers seemed to be much less engaged in the test.

There is no built-in way to interact with workers while they are doing the job on mTurk (though one can run an *external* HIT which one designs to run on one's own website; while this can require substantially more work, one can program any functionality one wants to have). But assuming we are talking about *internal* HITs on mTurk, we cannot provide further instructions to workers in real-time if they are unclear about any of the tasks or questions (they can send email). Similarly, there is no way to ask participants to "think out loud" while they are performing the tasks. If there is anything unclear or interesting in their feedback, it is very difficult to ask participants to elaborate.

mTurk workers seem unlikely to spend time giving substantial feedback to open-ended questions. A few words or a sentence is the most likely response to any open-ended questions. Deriving useful feedback from such answers for usability design can be quite challenging.

Specific user groups are difficult to identify. Such specific user groups may be unlikely to have a useful presence at present among online crowd workers. For instance, users with low computer literacy are unlikely to have an account with an online crowdsource platform like mTurk or uTest.

Spamming also appears to be a widespread phenomenon on mTurk. Because the purpose of usability testing is to find problems users may have, or mistakes they may make when using a website, it can be challenging to define good Gold Unit questions to detect spammers. While one can manually look at participant responses to detect cheating, this is far from ideal. Such a manual process reduces the benefit which is one of the main motivations for employing crowdsourcing in the first place.

## 6. FUTURE WORK
The results and analysis we have presented are based on a lab usability test and a crowdsourced usability test that were performed on the same website. While similar, the tests were not identical. As mentioned earlier, we utilized different tasks due to some known differential limitations of the two testing methods (e.g., difficulty of finding users experienced with the target web site in the crowdsourced environment). Our goal was not to perform an experiment with "type of usability test" as the independent variable. Nonetheless, we believe an important direction for future research is to explore how to conduct more parallel studies in the face of such crowdsourcing challenges.

For us, the different participants available via each method necessitated difference in the tasks each group performed. Lab test participants were all experienced users of the website with a better understanding of its content and structure. The usability problems identified by them were not only due to the experience of the website during the test, but also from previous experience. None of the crowdsourcing test participants had any experience with the website. It is likely that the problems they identified were more learnability problems than day-to-day usability problems. Due to the difference in familiarity of the two participant groups with the target web site, they were given different tasks to perform. Some usability problems found in some tasks may not be evident in other tasks.

While to some extent the demographic differences in participants were driven by core differences in methods being tested, measures could be developed to facilitate more parallel study. For example, our lab usability could be done with participants who do not have previous experience with the website, similar to crowd workers. On the other hand, we could also try to recruit students to use the online crowdsourcing platform. Either way, participants in both settings could be expected to perform maximally similar tasks to ensure a more systematic comparison.

Time-on-task and the actions on the website were not monitored in our crowdsourced usability test, though time-on-task can be very helpful in identifying usability problems. This information can be monitored rather than relying on self-reported data from workers by accessing the log data on the server end of the website. In this way, it would be possible to more accurately measure time-on-task and what unnecessary steps participants take before getting to the information they are looking for. Of course, this would clearly make for a more labor-intensive test.

As noted earlier, it is not possible to reject results or offer bonuses to workers via CrowdFlower. Workers who gave substantial feedback, workers who gave minimal feedback, and spammers all received the same compensation in the final test. A future test might explore a tiered payment system, rewarding higher quality workers with greater payment [6].

Another possible way to collect substantial feedback is to require a certain length of the answers to open-ended questions or a certain time spent on a question. For example, one could use javascript to disable or hide the "submit" button until workers input at least 30 words in the answer box or spend at least a minute working on an open-ended question. We could continue to refine the testing interface (rather than the website being tested) to make it more effortful to cheat rather than engage [17].

At some stages of development, companies may not need to set many requirements for its usability testers, especially when testing websites designed for use by the general public. Such cases would seem to be especially good candidates for crowdsourced usability testing on platforms like mTurk.

## 7. CONCLUSION

This paper explored crowdsourcing as an alternative way to conduct remote usability testing. We performed a lab usability test and a crowdsourced usability test on a graduate school's website. We found that while quality of results from crowdsourcing were typically not as good as those from our lab usability testing, some important usability problems can be identified via crowdsourced usability tests. Crowdsourcing appears to live up to its reputation of being faster, cheaper, and easier to perform with participants from diverse backgrounds. We believe crowdsourcing can be a useful tool for some usability testing scenarios, especially for those design/development teams who have only limited time and money. However, getting useful results and minimizing spam requires careful design of tasks and surveys. Crowdsourcing reduces implementation barriers but still requires careful experimental design and controls, and it introduces some new risks to be carefully managed.

We compared a single round of each type of test looking at a largely static website. While crowdsourcing tapped into a larger pool of respondents, feedback received from crowd workers was both shorter and less useful than that from lab test participants, often the difference being that between a single phrase and a page or more of transcripts. However, usability testing is intended to be done in multiple rounds, performed frequently over the course of developing and changing a design [2]. Because the crowdsourced testing costs appear to be so low relative to lab testing costs, an organization that could traditionally afford only one or two rounds of lab testing might potentially afford orders of magnitude more crowdsourced tests. The cumulative results of crowdsourcing may well be of greater value to an organization than a smaller number of lab tests. Certain metrics (e.g., time-on-task) and the identification of certain types of problems (e.g., those for users with previous experience on a predecessor web site) may be best associated with traditional lab testing. A hybrid test plan, involving both traditional and crowdsourced testing, may be the best solution for an emerging web site or application design. Arriving at a more nuanced understanding of relative return on investment of traditional vs. crowdsourced usability testing will be an important direction of future work for the field.

## 8. ACKNOWLEDGMENTS

This work was partially supported by a $25 educational gift from CrowdFlower. Any opinions, findings, conclusions or recommendations expressed in this material are those of the authors and do not necessarily reflect the views of their institution or funding agencies.

## 9. REFERENCES

[1] Amazon 2011. Amazon Mechanical Turk. http://aws.amazon.com/mturk/#pricing. Visited Sept. 2011.

[2] Bailey, G.D. 1993. Iterative methodology and designer training in human-computer interface design, *INTERCHI '93*, 198-205.

[3] Bias, R. G., & Huang, S. C. 2010. Remote, remote, remote, remote usability testing. Proceedings of the Society for Technical Communication Summit, May, Dallas.

[4] Bias, R. G., & Mayhew, D. J. (Eds.) 2005. Cost-justifying usability, 2nd edition: Update for the Internet age. San Francisco: Morgan Kaufmann.

[5] Caraway, B., 2010. Online labour markets: an inquiry into oDesk providers. In *Work Organizations, Labour and Globalisation. Vol. 4, Number 2.* Autumn, 2010 (111-125).

[6] Chen, D., & Dolan, W. 2011. Collecting Highly Parallel Data for Paraphrase Evaluation. In *Proc. of the Association for Computational Linguistics* (ACL).

[7] CrowdFlower 2011. FAQ - Self-Service – CrowdFlower. http://crowdflower.com/self-service/faq. Visited September 2011.

[8] Downs, J. S., Holbrook, M. B., Sheng, S., & Cranor, L. F. 2010 Are your participants gaming the system? Screening Mechanical Turk Workers. In *CHI '10: Proc. of the 28th international conference on Human factors in computing systems, ACM*, pp. 2399–2402. 2010

[9] Frei, B., 2009 "Paid crowdsourcing: Current state & progress toward mainstream business use," *Whitepaper Produced by Smartsheet.com*, 2009.

[10] Howe, J., Crowdsourcing: A Definition. http://crowdsourcing.typepad.com. Visited Sept. 2011.

[11] Howe, J. 2008. Crowdsourcing: Now with a real business model! http://www.wired.com/epicenter/2008/12/crowdsourcing-n. Visited September 2011.

[12] Ipeirotis, P. 2010a. Demographics of Mechanical Turk. *CeDER-10–01 working paper*, New York University.

[13] Ipeirotis, P. 2010b Be a Top Mechanical Turk Worker: You Need $5 and 5 Minutes. http://behind-the-enemy-lines.blogspot.com/2010/10/be-top-mechanical-turk-worker-you-need.html. Visited September 2010.

[14] Ipeirotis, P., Provost, F., & Wang, J. (2010) Quality Management on Amazon Mechanical Turk. In *HCOMP '10*, New York, NY, USA, 2010.

[15] ISO 9241-11 Ergonomic requirements for office work with visual display teminals (VDT)s – Part 11 Guidance on Usability, International Standard, 1998

[16] Kapelner, A., & Chandler, D. 2010. Preventing Satisficing in Online Surveys: A "Kapcha" to Ensure Higher Quality Data. In *CrowdConf Proceedings,* 2010.

[17] Kittur, A., Chi, E. H., & Suh, B. 2008. Crowdsourcing user studies with Mechanical Turk. In the *Proceedings of the ACM CHI Conference*.

[18] Kochhar, S., Mazzocchi, S., & Paritosh P. (2010) The anatomy of a large-scale human computation engine. In HCOMP '10, New York, NY, USA, 2010.

[19] Mason, W. and Watts, D.J. 2010. Financial incentives and the performance of crowds. In *ACM SIGKDD Explorations Newsletter* v. 11 no. 2, pp.100-108, 2010.

[20] Mason, W. and Suri, S. 2010. A Guide to Conducting Behavioral Research on Amazon's Mechanical Turk. In *Social Science Research Network*, pre-print. 2010.

[21] Molich, R., & Dumas, J. 2008. Comparative usability evaluation (CUE-4). *Journal of Behaviour & Information Technology, 27,* 3, 263-281.

[22] Nielsen, J. 2003. Usability 101: Introduction to Usability. http://www.useit.com/alertbox/20030825.html. Visited September 2011.

[23] Nielsen, J., & Mack, R. L. (Eds.) 1994. *Usability inspection methods*. New York: Wiley and Sons, Inc.

[24] Quinn, A. J., & Bederson, B. B., 2011 Human Computation: A Survey and Taxonomy of a Growing Field. In *Proceedings of CHI 2011,* May 7-12, 2011, Vancouver, BC, Canada.

[25] Ross, J., Irani, I., Silberman, M. Six, Zaldivar, A., & Tomlinson, B. 2010. "Who are the Crowdworkers?: Shifting Demographics in Amazon Mechanical Turk". In*: CHI EA 2010.* (2863-2872).

[26] Rubin, J., & Chisnell, D., 2008. *Handbook of usability testing: How to plan, design, and conduct effective tests.* Hoboken, NJ: Wiley.

[27] Spool, J., & Schroeder, W. 2001. Test web sites: five users is nowhere enough. In *Proc. CHI2001 Extended Abstract*. Seattle: ACM Press. 285-286

[28] uTest 2011a. About Us. http://www.utest.com/company/. Visited September 2011.

[29] uTest 2011b. How It Works. http://www.utest.com/how-it-works/agile-testing/. Visited Sept., 2011.

[30] uTest 2011c. Pricing Information. http://www.utest.com /pricing. Visited Sept. 2011.

[31] Winsor, J. 2009. Crowdsourcing: What it means for innovation. http://www.businessweek.com/innovate/content/jun2009/id20090615_946326.htm. Visited Sept. 2011.